\documentclass[prd,eqsecnum,notitlepage,%linenumbers,
nofootinbib,superscriptaddress]
{revtex4-1}
\global\arraycolsep=2pt
\usepackage{stmaryrd}
\usepackage{amsmath}
\usepackage{amssymb}
\usepackage{graphicx}
\usepackage{textcomp}
\usepackage{calrsfs}
\usepackage{yfonts}
\usepackage{bm}
\usepackage{soul}
\usepackage[multiple]{footmisc}
\usepackage{color}
\newcommand{\ben}{\begin{equation*}}
\newcommand{\een}{\end{equation*}}
\newcommand{\bean}{\begin{eqnarray*}}
\newcommand{\eean}{\end{eqnarray*}}

\newcommand{\nn}{\nonumber}
\newcommand{\be}{\begin{equation}}
\newcommand{\ee}{\end{equation}}
\newcommand{\bea}{\begin{eqnarray}}
\newcommand{\eea}{\end{eqnarray}}

\DeclareMathOperator{\tr}{tr}
\DeclareMathOperator{\csch}{csch}

\DeclareMathOperator{\sgn}{sgn}
\DeclareMathOperator{\re}{Re}
\DeclareMathOperator{\im}{Im}

\begin{document}
\title{Vacuum torque, propulsive forces, and anomalous tangential forces:
Effects of nonreciprocal media out of thermal equilibrium}
\author{Kimball A. Milton}
  \email{kmilton@ou.edu}
  \affiliation{H. L. Dodge Department of Physics and Astronomy,
University of Oklahoma, Norman, OK 73019, USA}
%\author{Prachi Parashar} 
%  \email{Prachi.Parashar@jalc.edu}
%  \affiliation{John A. Logan College, Carterville, IL
%62918, USA}
%\affiliation{Department of Energy and Process Engineering,
%Norwegian University of Science and Technology, 7491 Trondheim, Norway}
\author{Xin  Guo}
\email{guoxinmike@ou.edu}
\affiliation{H. L. Dodge Department of Physics and Astronomy, University of
Oklahoma, Norman, OK 73019, USA}
\author{Gerard Kennedy}\email{g.kennedy@soton.ac.uk}
\affiliation{School of Mathematical Sciences,
University of Southampton, Southampton, SO17 1BJ, UK}
\author{Nima Pourtolami}
\email{nima.pourtolami@gmail.com}
\affiliation{National Bank of Canada, Montreal, Quebec H3B 4S9, Canada}
\author{Dylan M. DelCol}
\email{dylan.m.delcol-1@ou.edu}
\affiliation{H. L. Dodge Department of Physics and Astronomy, University of
Oklahoma, Norman, OK 73019, USA}

\begin{abstract}From the generalized fluctuation-dissipation theorem, 
it is known that a body at rest made of  nonreciprocal material may experience 
a torque, even in vacuum, if it is not in thermal equilibrium with its 
environment. However, it does not experience self-propulsion in
such circumstances, except in higher order.  Nevertheless,
such a body may experience  both a normal torque and a  
lateral force when adjacent to  an
ordinary surface with transverse translational symmetry.  We explore how these
phenomena arise, discuss what terminal velocities might
be achieved, and point out  some of the limitations of
applying our results to observations, including the 
Lorenz-Lorentz correction, and the cooling due to radiation.
In spite of these limitations, the effects discussed would
seem to be observable.
\end{abstract} 

\date\today
\maketitle

\section{Introduction}
There is a long history of theoretical predictions of quantum or Casimir 
friction, where a particle or extended body that
moves parallel to a surface experiences a force opposing its motion.
The subject seems to have originated with Teodorovich \cite{teod} and
Levitov \cite{lev}.
For a selected bibliography on this subject before 2016, see Ref.~\cite{rev}.
The friction is typically conceived to arise because of dissipation in the 
surface. For a subset of papers on this subject, the reader is referred to
Refs.~\cite{pendry,volokitin,dedkov,barton,intravaia,davies}.
For a readable overview, see Ref.~\cite{marty}.
Quantum friction can also result
 if the body itself is made of dissipative material.
However, much earlier, it was recognized that, even in vacuum,
a moving body or an atom without dissipation will experience friction due 
to the surrounding radiation field---this is the
famous Einstein-Hopf effect \cite{EinsteinHopf}. 

In previous papers, we have considered quantum vacuum friction due to field
and dipole fluctuations \cite{guo,guo2}.  For low
velocities, the condition that the atom or particle  not gain or lose
energy, the Nonequilibrium Steady State condition (NESS)
\cite{reiche}, implies equal 
temperatures of the body and the environment, while relativistic velocities
typically imply that the body be substantially hotter than the environment. 
(For other earlier work on nonequilibrium
 friction, see Refs.~\cite{dedkov2,pieplow,vp}
for example.)  The forces
we considered there are true frictions, in that they always oppose the motion,
and they vanish at zero velocity and at zero temperature.

Here, we consider forces and torques that arise in the vacuum, or near other
bodies, when the relative velocity is zero.  This requires not only that the
system be out of equilibrium, so that the temperature of the body is
different from that of its environment, but 
also that the electrical properties that
characterize the material constituting the bodies be exotic, 
``nonreciprocal,'' at least in lowest order.
 Nonreciprocity seems not to be possible for an isolated
body; the typical way it can be achieved is through the introduction of
an external magnetic field, or some other appropriate external influence.
Thus, it is something of an oxymoron to discuss nonreciprocal vacuum torque
or friction.

Of course, there is much earlier work on such nonequilibrium phenomena,
involving heat transfer, torque, and nonreciprocal surface forces
\cite{krugeremig,Ott,maghrebi,Khandekar,khandekar0}. 
(Further references will be provided
as our discussion continues.)

The outline of this paper is as follows.  In Sec.~\ref{fdt} we discuss
how the fluctuation-dissipation theorem is modified for a nonreciprocal
susceptibility.  We then display, in Sec.~\ref{sec:model}, 
a simple model of such a nonreciprocal
material as a simplification of that given in Ref.~\cite{guofan}.
The corresponding 
quantum vacuum torque, first found in Refs.~\cite{fan,strekha}, 
is derived in Sec.~\ref{vactorque}.   In Sec.~\ref{sec:rot} we rederive
the modified torque if the body is slowly rotating, which was 
first worked out in Ref.~\cite{guofan}.
If the body is hotter or not too much colder than the environment, the
ordinary quantum frictional torque
 acts as a drag, and the body acquires a terminal
angular velocity which should be readily observable,
provided this temperature difference can be maintained.
In Sec.~\ref{sec:torquepl}, 
the effect on the torque of an underlying plate, be it
a dielectric slab or a perfect conductor, is investigated.  Then we turn
to the force, which can, of course, be inferred from the torque.
The quantum vacuum force is shown to vanish, in the weak-susceptibility
approximation that we use [Sec.~\ref{vacforce}], 
while, if an underlying surface is present, 
there is a component of the force parallel to the surface, 
as shown in examples of an imperfectly and a perfectly 
conducting plate [Sec.~\ref{transverseforce}]. Again, if
the nonequilibrium temperature difference could be maintained, a substantial
terminal velocity could be achieved.
In Sec.~\ref{sec:cool} we calculate the time it would take for a body
at rest to reach thermal equilibrium with the environment, 
unless some mechanism were 
supplied to keep it hotter or colder than the background.
Possible suppression effects due to the Lorenz-Lorentz 
(Clausius-Mossotti) correction are discussed in Sec.~\ref{sec:LL},
although the resulting torques and forces should
still be experimentally measurable.
Conclusions round out the paper.

Throughout, we use Heaviside-Lorentz (HL) electromagnetic units.  We also
set $\hbar=c=1$, except when numerical values are given.  
It should be noted that many authors (including some of us) often use
Gaussian (G) units, for which the polarizability 
differs by a factor of $4\pi$.

\section{Generalized fluctuation-dissipation theorem}
\label{fdt}
Let $\mathbf{x}(t)$ be some dynamical variable (such as an electric 
dipole moment).
In terms of its frequency Fourier transform, the fluctuation-dissipation
theorem (FDT) tells us the expectation value of the symmetrized 
quadratic product of the frequency-transformed variables:
\be
\langle S \mathbf{x}(\omega)\mathbf{x}(\nu)\rangle
=2\pi \delta(\omega+\nu)\Im \bm{\chi}(\omega)
\coth\frac{\beta\omega}2,
\ee
where $\beta=1/T$ is the inverse temperature of the system, and $\bm{\chi}$
is the generalized susceptibility; alternatively, the fluctuating quantity
might be the electric field, driven by the electric polarization, and
the susceptibility would be the retarded electric Green's function.

Typically, we regard the susceptibility tensor as diagonal, or at least 
symmetric.  More generally, the ``imaginary part'' that occurs in the FDT
is the anti-Hermitian part,
\be
\Im \bm{\chi}=\frac1{2i}(\bm{\chi}-\bm{\chi}^\dagger),\label{Imchi}
\ee
that is,
\be
(\Im\bm{\chi})_{ij}(\omega)=\frac1{2i}[\chi_{ij}(\omega)-\chi_{ji}^*(\omega)]
=\frac1{2i}[\chi_{ij}(\omega)-\chi_{ji}(-\omega)],
\ee
which uses the fact that $\chi_{ij}(\omega)$ is the Fourier transform of
a real response function. Unusual properties emerge from this if $\bm{\chi}$ 
is not symmetric:
This means that $\Im \bm{\chi}$ has both real and imaginary parts in the
conventional sense. The real part is
\begin{subequations}
\be
\re(\Im\bm{\chi})_{ij}(\omega)=\frac12[\im\chi_{ij}(\omega)+\im\chi_{ji}(\omega)],
\label{reIm}
\ee
which is symmetric in the indices but odd in $\omega$.  These are the
components that give rise to the quantum friction force and torque. The
imaginary part of $\Im\bm{\chi}$ is
\be
\im(\Im\bm{\chi})_{ij}(\omega)=-\frac12[\re\chi_{ij}(\omega)-\re\chi_{ji}(\omega)],
\label{imIm}
\ee
which is antisymmetric in the indices but even in $\omega$.
\end{subequations}
This property, as we shall see, leads to unusual phenomena for
nonreciprocal bodies, spontaneous quantum torque and quantum propulsion.
The term ``nonreciprocal'' seems to have a variety of meanings in the
literature; in this paper we will take it to mean that Eq.~(\ref{imIm}) is
nonzero.

For an ordinary material, $\chi_{ij}(\omega)$ is symmetric in the indices, 
which means that the anti-Hermitian part coincides with the usual imaginary
part:
\be
\chi_{ij}(\omega)=\chi_{ji}(\omega) \Rightarrow (\Im\bm{\chi})_{ij}(\omega)=
\im\chi_{ij}(\omega).
\ee
Where a susceptibility depends on continuous coordinates as well,
such as the Green's dyadic that describes the electric field, reciprocity
means invariance under interchange of discrete indices and continuous 
coordinates:
\be
\Gamma_{ij}(\mathbf{r,r'};\omega)=\Gamma_{ji}(\mathbf{r',r};\omega)\Rightarrow
(\Im \bm{\Gamma})_{ij}(\mathbf{r,r'};\omega)
=\im\Gamma_{ij}(\mathbf{r,r'};\omega).
\ee
It is easy to check that this is satisfied by the Green's dyadic for a 
dielectric half-space, for example, which has off-diagonal elements, symmetric 
in the tensor indices.
When the latter is expressed as a two-dimensional Fourier transform,
as is convenient when the environment consists of a dielectric slab
perpendicular to the $z$ axis, 
\be
\Gamma_{ij}(\mathbf{r,r'};\omega)=\int\frac{(d\mathbf{k}_\perp)}{(2\pi)^2}
e^{i\mathbf{k_\perp\cdot(r-r')}_\perp}g_{ij}(z,z';\mathbf{k}_\perp,\omega),
\label{fourierg}
\ee
the FDT is expressed in terms of 
\begin{equation}
(\Im \mathbf{g})_{ij}(z,z';\mathbf{k}_\perp,\omega)=\frac1{2i}\left[
g_{ij}(z,z';\mathbf{k}_\perp,\omega)-g_{ji}(z',z;-\mathbf{k}_\perp,-\omega)
\right].
\end{equation}

\section{Model for nonreciprocal material}
\label{sec:model}
In order to create a nonreciprocal response, one needs an external influence.
Such is supplied by a magnetic field.  Let us suppose a  oscillator 
with damping $\eta$ is
driven by both an electric and a magnetic field:
\be
m\frac{d^2\mathbf{r}}{dt^2}+m\eta\frac{d\mathbf{r}}{dt}
+m\omega_0^2\mathbf{r}=e\left(
\mathbf{E}+\frac{d\mathbf{r}}{dt}\times\mathbf{B}\right).
\ee
If the magnetic field lies in the $z$ direction, this immediately yields an
electric susceptibility that is nonsymmetric and nonreciprocal:
\be
\bm{\chi}=\omega_p^2\left(\begin{array}{ccc}
\frac{\omega_0^2-\omega^2-i\omega\eta}{(\omega_0^2-\omega^2-i\omega\eta)^2-
\omega^2\omega_c^2}&\frac{-i\omega\omega_c}{(\omega_0^2-\omega^2
-i\omega\eta)^2-\omega^2\omega_c^2}&0\\
\frac{i\omega\omega_c}{(\omega_0^2-\omega^2-i\omega\eta)^2
-\omega^2\omega_c^2}&\frac{\omega_0^2-\omega^2-i\omega\eta}{(\omega_0^2
-\omega^2-i\omega\eta)^2-\omega^2\omega_c^2}&0\\
0&0&\frac1{\omega_0^2-\omega^2-i\omega\eta}
\end{array}\right),\label{modforchi}
\ee
in terms of the plasma frequency $\omega_p^2=n e^2/m$, $n$ being the density
of charges, and the cyclotron frequency $\omega_c=eB/m$.  For a metal, we would
set the restoring force to zero, so $\omega_0=0$, and we exactly recover the
form given by Guo and Fan \cite{guofan}.
In particular, this provides us with a model for the anti-Hermitian part of
the susceptibility,
\be
\chi_{xy}=-\chi_{yx}=-i\frac{\omega_p^2\omega_c/\omega}{(\omega+i\eta)^2
-\omega_c^2}.
\ee
Numerically, for the charge and mass of the electron, 
\be
\omega_c=\frac{eB}m=m\frac{B}{B_c},\quad B_c=\frac{m^2}e=4.41 \times 10^9
\,\mbox{T},\ee
so for a magnetic field of strength  1\,T, $\omega_c\sim 10^{-4}$ eV, far
smaller than the damping parameter for gold, for example, $\eta\approx
3.5\times 10^{-2}$ eV. Thus, to a good approximation, we can use for 
a metal,
\be
\chi_{xy}-\chi_{yx}\approx- 2i\frac{\omega_c\omega_p^2}\omega\frac{1}{
(\omega+i\eta)^2}.\label{modela}
\ee

\section{Quantum vacuum torque}
\label{vactorque}
The vacuum torque on a stationary body  was discussed in 
general terms
in Ref.~\cite{fan}, and subsequently in Ref.~\cite{strekha}.
Somewhat earlier, Ref.~\cite{maghrebi} showed that a topologically insulating
film in a magnetic field, out of thermal equilibrium, experiences a torque,
which seems to be an instance of this same phenomenon.
A torque on a body at rest
requires that it be composed of nonreciprocal material, which is characterized
by having a real part of the susceptibility which is  nonsymmetric,
and that the temperature of the body, $T'$, be different from that of
the environment, $T$.

The torque on an arbitrary body, described by electric susceptibility,
$\bm{\chi}(\mathbf{r};\omega)$, is, in terms of the polarization, $\mathbf{P}$,
\begin{equation}
\bm{\tau}(t)=\int (d\mathbf{r})\,\mathbf{r}\times\left[-\bm{\nabla}\cdot 
\mathbf{P}(\mathbf{r},t)\mathbf{E}(\mathbf{r},t)
+\frac{\partial\mathbf{P}(\mathbf{r},t)}{\partial t}\times 
\mathbf{B}(\mathbf{r},t)\right].
\end{equation}
Writing this in terms of Fourier transforms, and eliminating $\mathbf{B}$ 
in favor of $\mathbf{E}$, we have
\bea
\bm{\tau}&=&\int (d\mathbf{r})\,\mathbf{r}\times\int
\frac{d\omega}{2\pi}\frac{d\nu}{2\pi}
e^{-i(\omega+\nu)t}\left\{-\bm{\nabla}\cdot\mathbf{P(r};\omega)
\mathbf{E(r};\nu)
-\frac\omega\nu\mathbf{P(r};\omega)\times[\bm{\nabla}\times\mathbf{E(r};\nu)]
\right\}\nonumber\\
&=&\int(d\mathbf{r})\frac{d\omega}{2\pi}\frac{d\nu}{2\pi}e^{-i(\omega+\nu)t}
\mathbf{r}\times\bigg[\frac\omega\nu \left\{\bm{\nabla}\cdot
[\mathbf{P(r;\omega)E(r;\nu)}]-\mathbf{P(r;\omega)}\cdot(\bm{\nabla})\cdot
\mathbf{E(r;\nu)}\right\}-\left(1+\frac\omega\nu\right)[\bm{\nabla}\cdot
\mathbf{P(r;\omega)]E(r;\nu)}\bigg],\nonumber\\\label{torque2}
\eea
where the notation in the 
first term in the last equality means that the $\bm{\nabla}$ is dotted only
with $\mathbf{P}$, but acts on both variables, while in the second term 
$\bm{\nabla}$ is the vector crossed with $\mathbf{r}$---the parentheses 
are intended to insulate it from the two vectors surrounding it.
Here, the source of the electric field is the electric polarization,
\begin{subequations}
\begin{equation}
\mathbf{E}(\mathbf{r};\omega)=\int (d\mathbf{r'})\,
\bm{\Gamma}(\mathbf{r,r'};\omega)\cdot
\mathbf{P}(\mathbf{r}';\omega),\label{EofP}
\end{equation}
where $\bm{\Gamma}$ is the retarded electromagnetic Green's dyadic,
while the polarization is linearly related to the electric field,
\begin{equation}
\mathbf{P}(\mathbf{r};\omega)=\int (d\mathbf{r})\,\delta(\mathbf{r-r'})
\bm{\chi}(\mathbf{r};\omega)\cdot \mathbf{E(r';\omega)}=
\bm{\chi}(\mathbf{r};\omega)\cdot\mathbf{E}(\mathbf{r};\omega),\label{PofE}
\end{equation}
\end{subequations}
where we assume that the electric susceptibility is local in space.  
This means that there are two origins for the quantum torque: 
field fluctuations and dipole fluctuations.
We evaluate the two contributions to the torque by use of the FDT:
\begin{subequations}
\label{fdtset}
\begin{eqnarray}
\langle S E_i(\mathbf{r};\omega)E_j(\mathbf{r}';\nu)\rangle&=&
2\pi\delta(\omega+\nu)(\Im\bm{\Gamma})_{ij}(\mathbf{r,r'};\omega)\coth
\frac{\beta\omega}2,\quad \beta=\frac1T,
\label{eefdt}\\
\langle S P_i(\mathbf{r};\omega)P_j(\mathbf{r}';\nu)
\rangle&=&2\pi\delta(\omega+\nu)\delta(\mathbf{r-r'})
(\Im\bm{\chi})_{ij}(\mathbf{r};\omega)\coth
\frac{\beta'\omega}2,\quad \beta'=\frac1{T'},\label{ppfdt}
\end{eqnarray}
\end{subequations}
where $S$ indicates that the symmetrized expectation values are used.
Therefore,  the last term in Eq.~(\ref{torque2}) vanishes, because
the sum of the two frequencies is zero, leaving us with\footnote{This
is made up of the ``internal'' and ``external'' torques, as given in 
Ref.~\cite{ce}, Eqs.~(4.47) and (4.46).}
\be
\bm{\tau}=\int(d\mathbf{r})\frac{d\omega}{2\pi}\frac{d\nu}{2\pi}
e^{-i(\omega+\nu)t}\left[
\mathbf{P(r;\omega)\times E(r;\nu)+P(r;\omega)\cdot(r\times\bm{\nabla})
\cdot E(r;\nu)}\right].\label{int+ext}
\ee
Here, the notation in the last term signifies that the free vector
index is on the angular momentum operator; the $\mathbf{P,E}$ are dotted
together.

Using Eqs.~(\ref{PofE}) and (\ref{eefdt}), 
we find for the EE contribution to the torque
\be
\bm{\tau}_i^{\rm EE}=
\int(d\mathbf{r})\frac{d\omega}{2\pi}\coth\frac{\beta\omega}2 \epsilon_{ijk}
\left[\chi_{jl}(\mathbf{r};\omega)(\Im\bm{\Gamma})_{lk}(\mathbf{r,r};\omega)
+\chi_{lm}(\mathbf{r};\omega)x_j\nabla'_k(\Im\bm{\Gamma})_{ml}(\mathbf{r,r'};\omega)
\bigg|_{\mathbf{r-r'\equiv R}\to\bm{0}}\right].\label{torque3}
\ee
%\textcolor{red}{(Check the relative sign---it propagates!)}
Here, $\bm{\Gamma}$ is taken to be the usual vacuum retarded Green's 
dyadic, $\bm{\Gamma}^0$,   the divergenceless part of which
 can be written as\footnote{The omitted term is proportional to 
$\delta(\mathbf{r-r'})$, which does not contribute in the point-split
\textit{limit}.}
\be
\bm{\Gamma}^{0\prime}(\mathbf{r,r'};\omega)=(\bm{\nabla\nabla-1}\nabla^2)
\frac{e^{i\omega R}}{4\pi R}
=\left[\bm{\hat R\hat R}(3-3i\omega R-\omega^2R^2)-\bm{1}(1-i\omega R
-\omega^2R^2)\right]\frac{e^{i\omega R}}{4\pi R^3},\label{gammaprime}
\ee
where $R=|\mathbf{r-r'}|$ and $\hat{\mathbf{R}}=\mathbf{R}/R$.
It is evident that the second term in Eq.~(\ref{torque3}) vanishes
here because $\im e^{i\omega R}/R$ is a function of $R$.
When (\ref{gammaprime})  is rotationally averaged in the coincidence limit 
($\mathbf{R}\to\bm{0}$),
we obtain 
\begin{equation}
    \bm{\Gamma}^{0\prime}(\mathbf{r,r'};\omega)
\to\bm{1}\left(\frac{\omega^2}{6\pi R}+i\frac{\omega^3}{6\pi}+O(R)\right).
\label{retg1}
\end{equation}
Therefore, we are left with only a single
term for the torque:
\be
\tau_i^{\rm EE}=\int\frac{d\omega}{2\pi} \coth\frac{\beta\omega}2\epsilon_{ijk}
\re\alpha_{jk}(\omega)\frac{\omega^3}{6\pi},
\ee
 where the mean polarizability\footnote{
The nonlinear effects occurring in the Lorenz-Lorentz law 
relate the polarizability to the permittivity; the polarizability is
implied thereby to be linear---see Ref.~\cite{entropy}, and 
Sec.~\ref{sec:LL}.} of the body\footnote{Note
that there is no requirement that the body be spherical, so any rotation
would be observable.} is given by
%so 
%Eq.~(\ref{meanpol}) pertains in the dilute limit.  \textcolor{red}{This is
%not particularly realistic, and such will be discussed further in 
%App.~\ref{appc}}}  
\begin{equation}
    \alpha_{jk}(\omega)=\int(d\mathbf{r})\chi_{jk}(\mathbf{r};\omega),
\label{meanpol}
\end{equation}
the real part of which is picked out by the necessity of the integrand being
even in $\omega$.
Thus, nonreciprocity is necessary for a vacuum torque in first
order.

For the PP fluctuation part, Eq.~(\ref{int+ext}) is written as
\be
\bm{\tau}^{\rm PP}=\int(d\mathbf{r})(d\mathbf{r'})\frac{d\omega}{2\pi}
\frac{d\nu}{2\pi}e^{-i(\omega+\nu)t}
\left[\mathbf{P(r};\omega)\times\bm{\Gamma}(\mathbf{r,r'};\nu)\cdot
\mathbf{P(r}';\nu)+\mathbf{P(r};\omega)\cdot(\mathbf{r}\times\bm{\nabla})
\cdot\bm{\Gamma}(\mathbf{r,r'};\nu)\cdot\mathbf{P(r'};\nu)\right],
\label{taupp}
\ee
to which the FDT (\ref{ppfdt}) is to be applied.
Again, for vacuum, the coincidence limit of the second term is zero, and
because of the diagonal form of the limit of the Green's dyadic, only
the antisymmetric part of the susceptibility survives upon use of 
Eq.~(\ref{ppfdt}).  That is, the $i$th component of the
quantity in square brackets in 
Eq.~(\ref{taupp}) becomes
\be
\delta(\mathbf{r-r'})2\pi\delta(\omega+\nu)
\epsilon_{ijk}(\Im\bm{\chi})_{jk}(\mathbf{r};\omega)
\left(-i\frac{\omega^3}{6\pi}\right)\coth\frac{\beta'\omega}2.
\ee
Here we have recognized that the antisymmetric part of 
$\Im\bm{\chi}$
%the anti-Hermitian
%part of the susceptiblity 
is even in $\omega$, according to Eq.~(\ref{imIm}),
and therefore only the odd part of the vacuum Green's dyadic survives.
The resulting torque is thus of the same form as for the EE contribution,
except for the sign, and the replacement $\beta\to\beta'$.
%while $\Im\bm{\chi}$ is the anti-Hermitian part of the susceptibility
%(\ref{Imchi}).
%\begin{equation}
%   \Im \bm{\chi}=\frac1{2i}(\bm{\chi}-\bm{\chi}^\dagger).
%\end{equation}
%The contribution to the torque 
%only comes from the antisymmetric part of the susceptibility, which is given by
%Eq.~(\ref{imIm}).
%\begin{equation}
%    \im\Im\chi_{ij}=-\frac12\re[\chi_{ij}(\omega)-\chi_{ji}(\omega)],
%\end{equation} which is even in $\omega$, as noted above.  
%The real part of $\Im\chi$ is symmetric and odd in $\omega$.

The combination of the two contributions thus yields
 the torque on a nonreciprocal body in 
vacuum, when the temperature of the body, $T'$, differs from that of the 
blackbody radiation, $T$, due to PP and EE fluctuations:
\begin{equation}
\tau_i=\int_{-\infty}^\infty \frac{d\omega}{2\pi} \frac{\omega^3}{6\pi}
\epsilon_{ijk}\re\alpha_{jk}(\omega)
\left[ \coth\frac{\beta\omega}2-\coth\frac{\beta'\omega}2\right].\label{vact}
    \end{equation}
This result exactly agrees with that of 
Guo and Fan \cite{fan}, for zero rotational velocity,  and with that of
 Strekha et al.~\cite{strekha}.  However, there is no quantum
vacuum force in this static situation, at least in first order,
 which we will demonstrate in Sec.~\ref{vacforce}. (Actually, this is already
evident from the vanishing of the external torque contribution.)

\begin{figure}
\includegraphics{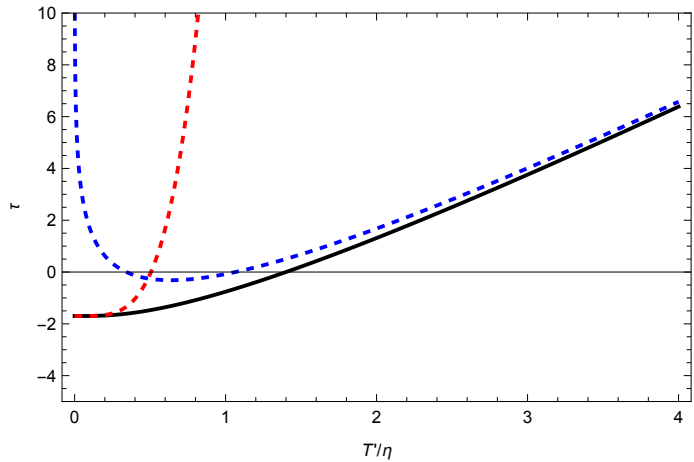}
\caption{\label{qvtfig}
The quantum vacuum torque, apart from the prefactor, in Eqs.~(\ref{tor0})
or (\ref{tor00}) is shown as a function of $T'/\eta$
by the solid line for $T/\eta=0.714$, appropriate for an environment at
room temperature, 300 K,
 and a gold body.  The dashed blue line is the
high-temperature
approximation (\ref{A9}), while the low-temperature approximation (\ref{A12})
 is shown by the dashed red  line. In both limiting
cases, the enviromental temperature
is treated exactly.}
\end{figure}

Let us use the model in Eq.~(\ref{modela}) to give an estimate of
the size of the torque.  Inserting this into Eq.~(\ref{vact})
and letting $\omega=\eta x$ gives
\be
\tau_z=\frac{\eta\omega_c\omega_p^2V}{3\pi^2}\int_{-\infty}^\infty
dx\frac{x^3}{(x^2+1)^2}\left(\coth\frac{\beta'\eta x}2
-\coth\frac{\beta\eta x}2\right)=\frac{4\eta\omega_c\omega_p^2V}{3\pi^2}
[I_2(\beta'\eta)-I_2(\beta\eta)],\label{tor0}
\ee
where $V$ is the volume of the body, and the integrals are defined
by Eq.~(\ref{A5}) of Appendix \ref{appa}.
As expected, this is positive if $T'>T$.
%If both $T/\eta$ and $T'/\eta$ are large, we could expand the 
%hyperbolic cotangents to get a linear temperature dependence:
%%\be
%%\tau_z\sim\frac{\omega_c\omega_p^2V}{3\pi}(T'-T),
%\quad T,T'\gg \eta.
%\label{torla}
%\ee
%However, for gold, for example, and the environment at room temperature,
%$T/\eta\sim 1$, so more relevant is an approximation valid for only $T'$ 
%large (see Appendix A for details):
%\be
%\tau_z\sim\frac{\eta\omega_c\omega_p^2 V}{3\pi^2}\left[\pi\frac{T'}\eta
%+2\ln\frac{2\pi T'}\eta+1+2\gamma_E-4\int_0^\infty\frac{x^3}{(x^2+1)^2}
%\frac1{e^{x\eta/T}-1}\right].\label{torlb}
%\ee
%Actually, it is straightforward to carry out the integrals in Eq.~(\ref{tor0})
%exactly, following Eq.~(B5) of Ref.~\cite{guo2}, in terms of digamma functions%,
%\be
%\int_0^\infty dx\frac{x^3}{(x^2+1)^2}\frac1{e^{\beta\eta x}-1}=\frac12
%\ln\frac{\beta\eta}{2\pi}+\frac14-\frac{\pi}{4\beta\eta}-\frac12
%\psi\left(\frac{\beta\eta}{2\pi}\right)-\frac{\beta\eta}{8\pi}\psi'
%\left(\frac{\beta\eta}{2\pi}\right),
%\ee
%although the integral form is just as convenient.
These integrals are readily evaluated in Eq.~(\ref{A6}):
\be
\tau_z=\frac{\eta\omega_c\omega_p^2V}{3\pi^2}\left[\frac{\pi}{\eta}(T-T')
+2\ln\frac{T'}T +2\psi\left(\frac\eta{2\pi T}\right)-2\psi\left(
\frac\eta{2\pi T'}\right)+\frac\eta{2\pi T}\psi'\left(\frac\eta{2\pi T}
\right)-\frac\eta{2\pi T'}\psi'\left(\frac\eta
{2\pi T'}\right)\right].\label{tor00}
\ee
The torque, and the approximations, are shown in Fig.~\ref{qvtfig}.
For a gold ($\omega_p=9$ eV, $\eta= 0.035$ eV)
 nanosphere of radius 100 nm, with $\omega_c=10^{-4}$ eV, the prefactor in
Eq.~(\ref{tor00}) is $8\times 10^{-25}$ Nm.
\section{Torque on a rotating body}
\label{sec:rot}
Of course, a torque on a body will cause it to rotate.  So, what is the
torque on a rotating body? Naturally, there should be a vacuum torque on a
rotating body made of ordinary (reciprocal) material, just as there is quantum 
vacuum friction on a linearly moving body \cite{guo,guo2}.  The nonreciprocal
aspect of this torque  was first treated in 
Ref.~\cite{fan}.

We consider a body rotating about the $z$ axis passing through its center of
mass with angular velocity $\Omega$.  The formula (\ref{int+ext}) should
still apply, with the external torque (the second term there) still not
contributing if the background is vacuum.  However,  the polarization and
electric fields should now refer to the body (rotating) frame, denoted
by a prime subsequently.  For low velocities, these are related to those
in the blackbody (unprimed) frame by a rotation:
\begin{subequations}
\bea
E_x'(\mathbf{r'},t)&=&E_x(\mathbf{r},t)\cos\Omega t+E_y(\mathbf{r},t)\sin
\Omega t,\\
E_y'(\mathbf{r'},t)&=&E_y(\mathbf{r},t)\cos\Omega t-E_x(\mathbf{r},t)\sin
\Omega t,
\eea
\end{subequations}
which means for the frequency transforms,
\begin{subequations}
\bea
E'_x(\mathbf{r}';\omega)&=&\frac12\left[E_x(\mathbf{r},\omega_+)
+E_x(\mathbf{r};\omega_-)\right]+\frac1{2i}\left[E_y(\mathbf{r};\omega_+)
-E_y(\mathbf{r};\omega_-)\right],\\
E'_y(\mathbf{r}';\omega)&=&\frac12\left[E_y(\mathbf{r},\omega_+)
+E_y(\mathbf{r};\omega_-)\right]-\frac1{2i}\left[E_x(\mathbf{r};\omega_+)
-E_x(\mathbf{r};\omega_-)\right],
\eea
\end{subequations}
where $\omega_\pm=\omega\pm\Omega$.  $\mathbf{P}$ transforms in the same way.

The strategy followed to calculate  the quantum rotational friction is 
the same as that used for quantum rectilinear friction \cite{guo2}.  
There are two contributions: field fluctuations and dipole fluctuations.  For
the former we use Eq.~(\ref{PofE}) to replace the polarization by the electric
field, but now understood in the body frame.
Then we have to transform both electric 
fields to the blackbody frame.  For vacuum
friction we can use Eq.~(\ref{retg1}) for the Green's dyadic that appears
when the FDT is employed for the fields.  We also only keep half the terms:
only those proportional to $\delta(\omega_\pm+\nu_\mp)$ do not average 
to zero in time. The result of a straightforward calculation is
\be
\tau_z^{\rm EE}=\int\frac{d\omega}{2\pi}\frac{\omega^3}
{6\pi}\left[\im(\alpha_{xx}+\alpha_{yy})(\omega_-)+\re(\alpha_{xy}
-\alpha_{yx})
(\omega_-)\right]\coth\frac{\beta\omega}2,
\ee
where we have also noticed that under $\omega\to-\omega$, $\omega_+\to
-\omega_-$.

The procedure for the PP fluctuations is similar, except now we replace 
$\mathbf{E}$ by $\mathbf{P}$ according to Eq.~(\ref{EofP}). 
This holds in the blackbody frame,
so $\mathbf{P}$ must be transformed back to the body (rotating) frame.  
Simplifications
occur as before for the vacuum case, and we find after a bit of algebra
\be
\tau_z^{\rm PP}=-\int\frac{d\omega}{2\pi}\frac{\omega^3}{6\pi}
\left[\im(\alpha_{xx}+\alpha_{yy})(\omega_-)+\re(\alpha_{xy}
-\alpha_{yx})(\omega_-)
\right]
\coth\frac{\beta'\omega_-}2.
\ee
Thus, when the two contributions are added, we find for the torque on a
(slowly) rotating body:
\be
\tau_z=\frac1{12\pi^2}\int_{-\infty}^\infty d\omega\,\omega_+^3\left[
\im(\alpha_{xx}+\alpha_{yy})(\omega)+\re(\alpha_{xy}-\alpha_{yx})(\omega)
\right]\left(\coth\frac{\beta\omega_+}2-\coth\frac{\beta'\omega}2\right).
\ee
This is precisely the torque found in Ref.~\cite{fan} (recall $4\pi \alpha^G
=\alpha^{\rm HL}$).  This result for an isotropic  (reciprocal) particle
was given in Ref.~\cite{pan}, which further considered the effect of a magnetic
field.\footnote{Earlier related works on forces and torques on 
bodiess with various kinds of asymmetries include Refs.~\cite{Reid, Ott, 
maghrebi, Khandekar}.}
Note that if $\Omega=0$ the first term involving the diagonal polarizabilities
vanishes because the integrand is odd, and the second term reproduces
Eq.~(\ref{vact}).

It is illuminating to expand this expression to leading order in the 
rotational velocity $\Omega$ (this is the adiabatic approximation):
\bea
\tau_z&=&\frac1{12\pi^2}\int_{-\infty}^\infty d\omega\,\omega^3\bigg\{
\re\left(\alpha_{xy}-\alpha_{yx}\right)(\omega)\left[\coth\frac{\beta\omega}2
-\coth\frac{\beta'\omega}2\right]\nn\\
&&\quad\mbox{}-\Omega\frac{3}{\omega}\im\left(\alpha_{xx}
+\alpha_{yy}\right)
(\omega)\left[\coth\frac{\beta'\omega}2-\coth\frac{\beta\omega}2\right]
-\Omega\frac{\beta}2\im\left(\alpha_{xx}+\alpha_{yy}\right)(\omega)\csch^2
\frac{\beta\omega}2\bigg\}.\label{torqueexp}
\eea
The first term here is the (nonreciprocal) quantum vacuum torque (\ref{vact}), 
the second is the
nonequilibrium contribution to the ordinary (reciprocal)
quantum vacuum frictional torque, 
and the third term is the analog of the Einstein-Hopf quantum vacuum friction.
The sum of the two frictional terms is a drag if $T'>T$. 
 If $T'<T$, the angular velocity changes sign, so initially
the friction remains a drag, but for sufficiently low temperatures, $T'$, the
second term in Eq.~(\ref{torqueexp}) will dominate, and exponential growth
of the angular velocity will ensue, insofar as the low-velocity approximation
remains valid.

Note, that if the last two terms constitute a drag, the nonreciprocal torque
found here will lead to the body rotating with a constant 
terminal angular velocity.  Writing Eq.~(\ref{torqueexp}) in the abbreviated
form,\footnote{That is, $\tau_0=\tau_z(\Omega=0)$ and 
$\tau_1'=-\frac{d\tau_z}{d\Omega}(\Omega=0)$.}
\be
\tau_z=I\dot{\Omega}=\tau_0-\Omega\tau'_1,
\ee
where $I$ is the moment of inertia of the body,  
we immediately obtain
\be
\Omega(t)=\frac{\tau_0}{\tau'_1}\left(1-e^{-\tau'_1 t/I}\right),
\ee
if the body is not rotating at time $t=0$.
The terminal
velocity is $\Omega_T=\Omega(t\to\infty)=\tau_0/\tau'_1$, which might
be expected to be small.  (This, of course, assumes that the particle and
environmental temperatures do not change.  We will address the tendency
toward thermal equilibrium in Sec.~\ref{sec:cool}.)

%The dissipative contribution to the torque is
%given by
%\be
%\tau_1'=\frac{1}{3\pi^2}
%\left\{\int_0^\infty d\omega\,\omega^3\frac{d\alpha_\perp}{d\omega}(\omega)
%\frac1{e^{\beta\omega}-1}+3\int_0^\infty d\omega\,\omega^2
%\alpha_\perp(\omega)\frac1{e^{\beta'\omega}-1}\right\},
%\quad\alpha_\perp(\omega)=\im(\alpha_{xx}+\alpha_{yy})(\omega). 
%\ee
To proceed, let us again use the model (\ref{modforchi}), with $\omega_0=0$
and $\omega_c\ll\eta$. Then we have
\be
\im(\alpha_{xx}+\alpha_{yy})(\omega)
=\frac{2\omega_p^2\eta}{\omega(\omega^2+\eta^2)}V.
%\quad
%\frac{d\alpha_\perp}{d\omega}(\omega)=-\frac{2\omega_p^2\eta(3\omega^2
%+\eta^2)}{\omega^2(\omega^2+\eta^2)^2}V.
\ee
The result is ($x=\omega/\eta$)
\bea
\tau_1'&=&
\frac{2\omega_p^2\eta V}{3\pi^2}\left(3+\beta\frac\partial{\partial\beta}
\right)\int_0^\infty dx\frac{x}{(x^2+1)}
\left[\frac1{e^{\beta'\eta x}-1}-\frac1{e^{\beta\eta x}-1}\right]
%+\frac23
%\int_0^\infty \frac{dx\,x}{(x^2+1)^2}\frac{1}{e^{\beta\eta x}-1}\right\}
\nonumber\\
&=&\frac{\omega_p^2\eta V}{3\pi^2}\left\{\frac{\pi}{\eta}(2T-3T')
+3\ln\frac{T}{T'}-1+3\psi\left(\frac\eta{2\pi T}\right)
-3\psi\left(\frac\eta{2\pi T'}\right)
+\frac\eta{2\pi T}\psi'\left(\frac\eta{2\pi T}\right)\right\}.
\label{tau1}
\eea
%From the first line of this, it is evident that $\tau'_1>0$ if $T'>T$, that 
%is, the torque is a drag if the particle is hotter than the environment.

\begin{figure}
%\begin{minipage}{.49\linewidth}
%\includegraphics{fig2a.eps}
%\end{minipage}\hfill
%\begin{minipage}{.49\linewidth}
%\includegraphics{fig2b.eps}
%\end{minipage}
\includegraphics{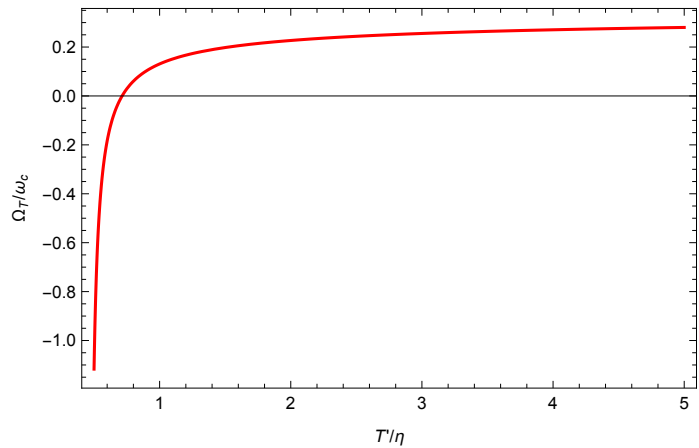}
\caption{\label{fig:omegaT} The terminal angular velocity of the nonreciprocal
body when it is  hotter or colder,  than its environment, 
which is taken to be at room 
temperature.  The limit of $\Omega_T$ for high 
particle temperature is universally $\omega_c/3$, independent of the
background temperature.
%The limit for low temperature, shown in the right panel, depends on the
%environmental temperature.}
When the temperature is lower than that of the background, the frictional
torque initially acts as a drag, but for sufficiently low temperature, 
the ``frictional'' terms change sign, and the angular velocity increases
exponentially  without bound.}
\end{figure}

Let us write, from Eqs.~(\ref{tor00}) and (\ref{tau1}),
\be
\tau_0=\frac{\eta\omega_c\omega_p^2V}{\pi^2}f(T,T'),\quad
\tau'_1=\frac{\eta\omega_p^2V}{\pi^2}g(T,T').
\ee
Then, the terminal angular velocity is
\be
\Omega_T=\frac{\tau_0}{\tau'_1}={\omega_c}\frac{f(T,T')}{g(T,T')}\sim
\omega_c\sim 10^{-4} \mbox{eV}\sim 10^{11}\, \mbox{s}^{-1},\label{term1}
\ee
perhaps surprisingly high, but very small compared to atomic frequencies.
(The terminal circumferential speed
in this case is $\Omega_T R\sim10^4$ m/s, for a gold 
nanosphere of radius $R=100$ nm.)
The relaxation time required to reach this velocity is 
\be
t_0=\frac{I}{\tau_1'}\sim \frac{M R^2}{\omega_p^2\eta V}\sim 10^6\, \mbox{s},
\label{relax1}
\ee
for the same parameters. The temperature dependence
of the terminal angular velocity, $\Omega_T$, 
is shown in Fig.~\ref{fig:omegaT}.
Note that if the temperature of the body 
is lower than that of the environment, the terminal
 angular velocity is negative.  Whether the body is
hotter or not too much  colder than the environment, it will  reach a terminal
velocity if the temperature difference is maintained.  If the temperature
of the body is much colder than that of the environment 
(for the example
shown, about half room temperature), the frictional torques reverse sign,
and no bound to the  angular
velocity can be reached. A negative $\tau'_1$ means exponential growth.
 Of course, before the angular velocity gets too
large, the nonrelativistic approximation used here
breaks down, even if the temperature difference
can be  maintained by some external or internal  agent.

However, it is not necessary to wait a long time to reach
the terminal velocity, because the initial angular acceleration,
\be
\dot\Omega(0)=\frac{\Omega_T}{t_0}\sim 10^5 \,\mbox{s}^{-2},\label{Omega0}
\ee
should be easily discernible.

\section{Torque in presence of a dielectric plate}
\label{sec:torquepl}
What happens if the background is less trivial, say consisting of an
isotropic dielectric
plate filling the halfspace $z<0$, while the body lies a distance $a$ above it?
Then, of course, there will be a torque on a nonspherical
 body as well as a force,
due to ordinary Casimir forces, even when the body is made of ordinary
reciprocal material.  What would be unusual is if there were a torque around
the $z$ axis, since the environment possesses rotational symmetry about
that direction. For simplicity, we will assume that the entire background,
vacuum plus dielectric plate, is in equilibrium at temperature $T$, while
the nanoparticle has temperature $T'$.
 In that case, the $z$-component of the torque coming
from field fluctuations is given by Eq.~(\ref{torque3}).
%\be
%\tau^{\rm{EE}}_z=\int (d\mathbf{r})\frac{d\omega}{2\pi}
% \epsilon_{3jk}\left[\chi_{jl}(\mathbf{r};\omega)
%\Im \Gamma_{lk}(\mathbf{r,r};-\omega)-\chi_{lm}(\mathbf{r};\omega)
%x_j\nabla'_k\Im\Gamma_{ml}(\mathbf{r,r'};-\omega)\right]\coth\frac{\beta
%\omega}2.\label{tauzEE}
%\ee
Now both terms can contribute, but only for a nonreciprocal
body, which might be irregularly shaped.  For such an body, where 
\be
\hat\chi_{ij}(\mathbf{r};\omega)=\re(\chi_{ij}-\chi_{ji})(\mathbf{r};\omega)
\ee
is nonzero, the normal torque component can be computed 
from the explicit construction of the Green's dyadic (for example,
as given in Ref.~\cite{guo}).  Using the Fourier representation 
(\ref{fourierg}), the integration over $k_x$ and $k_y$
will vanish except for the $g_{xx}$ and $g_{yy}$ terms, for the first
term in Eq.~(\ref{torque3}), while $g_{xz,zx}$ and $g_{yz,zy}$ contribute
for the second term, yielding the scattering part of the torque:
\bea
\tau_z^s&=&\int (d\mathbf{r})\frac{d\omega}{2\pi}
\left[\coth\frac{\beta\omega}2-\coth\frac{\beta'\omega}2\right]
\frac14\int\frac{(d\mathbf{k}_\perp)}{(2\pi)^2}
\bigg\{\hat\chi_{xy}(\mathbf{r};\omega)
\im\left[\left(\kappa r^H+\frac{\omega^2}{\kappa}
r^E\right)e^{-2\kappa z}\right]\nn\\
&&\quad\mbox{}+[\hat\chi_{yz}(\mathbf{r};\omega)x-
\hat\chi_{xz}(\mathbf{r};\omega)y] k^2\im\left[r^H e^{-2\kappa z}
\right]\bigg\},\label{torqueback}
\eea
where $\kappa=\sqrt{k^2-\omega^2}$, and the transverse magnetic 
and electric reflection coefficients are
\be
r^H=\frac{\kappa-\kappa'/\varepsilon(\omega)}{\kappa
+\kappa'/\varepsilon(\omega)},
\quad
r^E=\frac{\kappa-\kappa'}{\kappa+\kappa'},
\quad \kappa'=\sqrt{k^2-\omega^2\varepsilon(\omega)}.
\ee
Now the torque depends on the distribution of the anisotropic material
across the body, so is not describable by simply an effective nonreciprocal
polarizability.

Note further that the second two terms in Eq.~(\ref{torqueback}), proportional
to $x$ and $y$, respectively, depend on the position of the body as well
as the distribution of material within the body.  If we write $\mathbf{r=
R+r'}$, where $\mathbf{R}$ locates the center of mass of the body, we can
read off the force on the center of mass of the body from $\tau_z=XF_y-YF_x$,
so that 
\be
F_x=\int(d\mathbf{r})\int_0^\infty\frac{d\omega}{2\pi}\left[\frac1{e^{\beta
\omega}-1}-\frac1{e^{\beta'\omega}-1}\right]\hat\chi_{xz}(\mathbf{r};\omega)
\int\frac{(d\mathbf{k}_\perp)}{(2\pi)^2}k^2\im\left[r^H e^{-2\kappa z}\right].
\label{forcefromtorque}
\ee
We will derive this result directly in Sec.~\ref{transverseforce}.

\subsection{Torque for a nanoparticle above a perfectly conducting plate}
A very simple example is provided by a perfectly conducting surface lying in
the $z=0$ plane. This means $r^{H,E}=\pm1$. 
 Consider only the case with $\hat\chi_{xy}\ne0$, that is, for our model,
the magnetic field lying in the $z$ direction.  The imaginary part comes only
from the region where $\omega^2>k^2$, where
\be
\im\kappa=-\sgn(\omega)\sqrt{\omega^2-k^2},\ee
 and then the integral in Eq.~(\ref{torqueback}) over transverse wavenumbers is
(provided the body is of negligible extent, a nanoparticle,
so $z=a$)
\bea
\frac1{(2\pi)^2}\int_0^{|\omega|} dk\,k\int_0^{2\pi}d\theta\left[-\sqrt{
\omega^2-k^2}-\frac{\omega^2}{\sqrt{\omega^2-k^2}}\right]
\cos\left[2a\sqrt{\omega^2-k^2}\right]
&=&-\frac{\omega^3}{2\pi}\int_0^1 dy(y^2+1)\cos 2\omega a y\nonumber\\
&=&-\frac1{\pi(2a)^2}[u\cos u+(u^2-1)\sin u].
\eea  Here we have defined $u=2\omega a$.  Then, we write the 
scattering part of the torque in the 
direction perpendicular to the plate as %\textcolor{red}{(again, ignoring
%nonlinear effects)}
\be
\tau^s_z=\frac2{\pi^2}V\omega_c\omega_p^2\eta \int_0^\infty du
\frac{ u\cos u+(u^2-1)\sin u}{(u^2+(2\eta a)^2)^2}
\left[\frac1{e^{\beta  u/(2a)}-1}-\frac1{e^{\beta' u/(2a)}-1}\right].
\ee
Close to the plate, $2\eta a\ll 1$, $\beta/(2a)\gg1$, $\beta'/(2a)\gg1$,
where the integral is dominated by small values of $u$, for which
the numerator in the integrand approaches
\be
 u\cos u+(u^2-1)\sin  u\sim \frac{2}3 u^3, 
\ee
we obtain  precisely the negative of the torque coming from the
vacuum contribution (\ref{tor0}).  That the total normal 
torque vanishes as the
perfectly conducting plate is approached is expected, because the tangential
electric field must vanish there.

\begin{figure}
\includegraphics{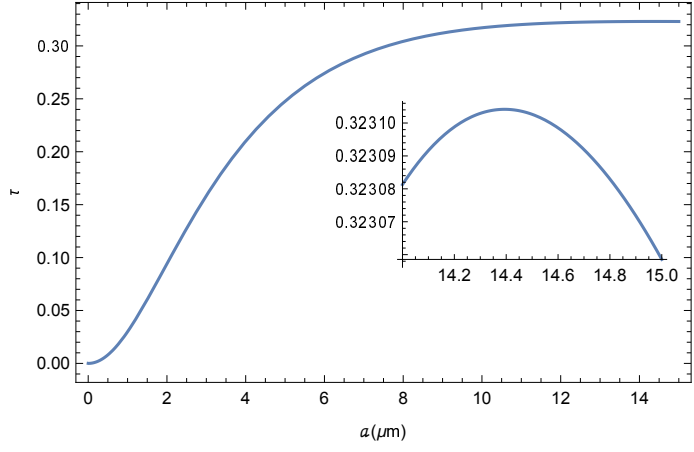}
\caption{\label{fig:torofa} Torque (apart from the prefactor in 
Eq.~(\ref{tottorpc})) as a function of separation of the 
nanoparticle $a$ from the perfectly conducting  plate in $\mu$m, for
a damping parameter of $\eta=0.035$ eV. The temperatures are taken to be
$T=300$ K and $T'=600$ K.  The torque vanishes, as expected, close to the
plate, and approaches the vacuum value far from the plate.  Most interesting
is the appearance of a very
 weak maximum at about 14.4 $\mu$m, just before the
decrease to the vacuum torque value, as displayed in the inset.}
\end{figure}

The total torque is then
\be
\tau^{\rm vac}_z+\tau_z^{\rm s}
=\frac4{3\pi^2}V\omega_c\omega_p^2\eta\int_0^\infty du 
\frac{u^3}{(u^2+(2a\eta)^2)^2}\left[1
-\frac3{2}\frac{ u\cos  u+(u^2-1)\sin  u}{ u^3}\right]
\left[\frac1{e^{\beta'  u/(2a)}-1}-\frac1{e^{\beta u/(2a)}-1}\right].
\label{tottorpc}
\ee
This is plotted in Fig.~\ref{fig:torofa}.

\section{Quantum vacuum force}
\label{vacforce}
Let us start by writing the force on a dielectric body on which an
electric field is impressed, writing the fields in terms of their
frequency transforms:
\bea
\mathbf{F}&=&\int(d\mathbf{r})\int \frac{d\omega}{2\pi}\frac{d\nu}{2\pi}
e^{-i(\omega+\nu)t}
\left\{-\left[\bm{\nabla}\cdot \mathbf{P}(\mathbf{r};\omega)\right]
\mathbf{E}(\mathbf{r};\nu)-i\omega\mathbf{P}(\mathbf{r};\omega)\times
\left[\frac{1}{i\nu}\bm{\nabla}\times \mathbf{E}(\mathbf{r};\nu)\right]
\right\}\nonumber\\
&=&\int(d\mathbf{r})\int \frac{d\omega}{2\pi}\frac{d\nu}{2\pi}
e^{-i(\omega+\nu)t}
\left\{-\left[\bm{\nabla}\cdot \mathbf{P}(\mathbf{r};\omega)\right]
\mathbf{E}(\mathbf{r};\nu)\left(1+\frac\omega\nu\right)
-\frac\omega\nu
\mathbf{P}(\mathbf{r};\omega)\cdot(\bm{\nabla})\cdot
 \mathbf{E}(\mathbf{r};\nu)\right\},
\eea
where in the second line we integrated spatially by parts.
Unlike for the torque, the total divergence does not contribute.
Now we either expand $\mathbf{P}$ in terms of $\mathbf{E}$, using
Eq.~(\ref{PofE}), or $\mathbf{E}$ in terms of $\mathbf{P}$, using
Eq.~(\ref{EofP}), and then use the fluctuation-dissipation theorem on the
two parts.   
This yields rather immediately for the force on the body\footnote{This
general formula can be derived immediately
from the torque on the center of mass inferred from Eq.~(\ref{torque3}).}
\be
F_i^{\rm EE}+F_i^{\rm PP}=\int\frac{d\omega}{2\pi}
(d\mathbf{r})\left[\chi_{jl}(\mathbf{r};\omega)
\nabla'_i(\Im\bm{\Gamma})_{lj}(\mathbf{r,r'};\omega)\big|_{\mathbf{r'=r}}
\coth\frac{\beta\omega}2+
(\Im\bm{\chi})_{jl}(\mathbf{r};\omega)\nabla_i
\Gamma_{lj}(\mathbf{r,r'};-\omega)\big|_{\mathbf{r'=r}}
\coth\frac{\beta'\omega}2\right].\label{transforce}
\ee
However, for vacuum,  Eq.~(\ref{retg1}) describes the vacuum Green's
dyadic, 
%Green's dyadic can be written as\footnote{Equation (\ref{retg1}) follows from 
%this.}
%\be
%\bm{\Gamma}^{0\prime}(\mathbf{r,r'};\omega)=(\bm{\nabla\nabla-1}\nabla^2)
%\frac{e^{i\omega R}}{4\pi R}
%=\left[\bm{\hat R\hat R}(3-3i\omega R-\omega^2R^2)-\bm{1}(1-i\omega R
%-\omega^2R^2)\right]\frac{e^{i\omega R}}{4\pi R^3},
%\ee
%where $\mathbf{R=r-r'}\to 0$.
So, again  in  the coincidence limit, it is then clear that
the gradient of the Green's dyadic vanishes, and thus there is no vacuum
force.  The conclusion appears to be  opposite to that of 
Refs.~\cite{muller,Reid}, but evidently the self-propulsion found there
arises as a second-order effect, with which we will deal in a later paper.
There, nonreciprocity is not required.  The only necessary conditions
are that the system be out of equilibrium, and that the body be 
extended and inhomogeneous.

\section{Transverse force on a nonreciprocal  nanoparticle
induced by  a dielectric surface}
\label{transverseforce}
In contrast to the result found in the previous section, 
a nonreciprocal body does experience, in first order, 
 a force transverse to another
ordinary body, even when both bodies are at rest, provided they are not in
thermal equilibrium with each other.  This was observed in Ref.~\cite{muller}
and more recently in Refs.~\cite{gelb,Khandekar}.

This is still described by the formula (\ref{transforce}), but requires
the scattering part of the Green's function.  We will consider the second,
ordinary body to be a planar dielectric,
with permittivity $\varepsilon(\omega)$, lying in the halfspace $z<0$, while
the nonreciprocal body lies at a distance $z=a$ above the plane.  It is
convenient then to introduce a two-dimensional Fourier transform in
the transverse coordinates, $x$ and $y$.  Then the force in the $x$ direction,
say, is
\be
F_x=-\int(d\mathbf{r})
\frac{d\omega}{2\pi}\frac{(d\mathbf{k}_\perp)}{(2\pi)^2}ik_x
\left[\chi_{jk}(\mathbf{r};\omega)
(\Im \mathbf{g}^s)_{kj}(z,z;\omega,\mathbf{k}_\perp)
\coth\frac{\beta\omega}2-(\Im\bm{\chi})_{jk}(\mathbf{r};\omega)
g^s_{kj}(z,z;-\omega,\mathbf{k}_\perp)\coth\frac{\beta'\omega}
2\right].
\ee
Here, the $s$ superscripts on the reduced Green's functions represent
the scattering parts, since it is evident that the bulk (vacuum) part
does not contribute, as already demonstrated in the previous section.
%From the explicit construction of the Green's dyadic (for example,
%as given in Ref.~\cite{guo})  
Now the integration over $k_x$ and $k_y$
will vanish except for $g_{xz}$ and $g_{zx}$.\footnote{For the bulk 
(vacuum) contribution, the force would involve the symmetric limit
$\lim_{z\to z'}\sgn(z-z')=0$.}
Hence, unlike for the torque,  only the TM
Green's function contributes. Using the properties of $\Im\bm{\chi}$ given in 
Sec.~\ref{fdt}, we immediately obtain
\be
F_x=2\int_0^\infty\frac{d\omega}{2\pi}\int\frac{(d\mathbf{k}_\perp)}{(2\pi)^2}
(d\mathbf{r})\hat\chi_{xz}(\mathbf{r};\omega)
k_x^2\im \left(r^H e^{-2\kappa z}\right)\left[\frac1{e^{\beta\omega}-1}-
\frac1{e^{\beta'\omega}-1}\right].
\ee
This force is precisely that inferred from the torque in 
Eq.~(\ref{forcefromtorque}).
%and we have adopted the abbreviation
%\be
%\xi(\omega)=\re[\alpha_{12}(\omega)-\alpha_{21}(\omega)], 
%\ee
%which would
%vanish for an ordinary material.

As with Casimir friction, this force will vanish unless dissipation 
occurs somewhere.  This  could be due to dissipation in 
the dielectric slab, or to radiation.  We will consider these in the following
subsections.
\subsection{Dissipation in a metallic slab}
%for example, in the dielectric slab.\footnote{More interesting, but more
%complicated, is the situation
%when the radiation itself provides the damping, through
%the appearance of an imaginary part in $\kappa$ and $\kappa'$}
We will  describe the metallic substrate  by  a Drude model,
\be
\varepsilon(\omega)=1-\frac{\omega^2_p}{\omega^2+i\omega\nu},
\ee
where $\omega_p$ is the plasma frequency, and $\nu$ the damping parameter.
For simplicity, we will  consider the regime
\be
\nu\ll\omega\ll\omega_p,\quad \omega\ll k, 
\ee
so that
\be \im \varepsilon(\omega)\approx \frac{\omega_p^2\nu}{\omega^3},
\ee
and then, for low frequencies,
\be
\im r^H=\im\frac{\kappa-\kappa'/\varepsilon}{\kappa+\kappa'/\varepsilon}
\approx \im\frac{\varepsilon-1}{\varepsilon+1}\approx
\frac{2\omega\nu}{\omega_p^2}.
\ee
%where $\kappa'=\sqrt{k^2-\omega^2\epsilon(\omega)}$.
Thus, in this approximation, where we crudely 
replace $\kappa$ and $\kappa'$ by $k$, the force on a nanoparticle
of negligible extent is %\textcolor{red}{(again, in the linear approximation)}
\bea
F_x&=&-2\frac{\nu}{\omega_p^2}\int\frac{d\omega}{2\pi}
\frac{(d\mathbf{k}_\perp)}{(2\pi)^2}\hat\alpha_{xz}(\omega)
\omega k_x^2 e^{-2\kappa a}\left(\frac1{e^{\beta\omega}-1}-\frac1{e^{\beta'
\omega}-1}\right)\nonumber\\
&\approx&-\frac{12}{(2\pi)^2}\frac1{(2a)^4}\frac\nu{\omega_p^2}
\int_0^\infty d\omega\,
\hat\alpha_{xz}(\omega)\omega\left(\frac1{e^{\beta\omega}-1}-\frac1{e^{\beta'
\omega}-1}\right).\label{lowfreqf}
\eea

\begin{figure}
\includegraphics{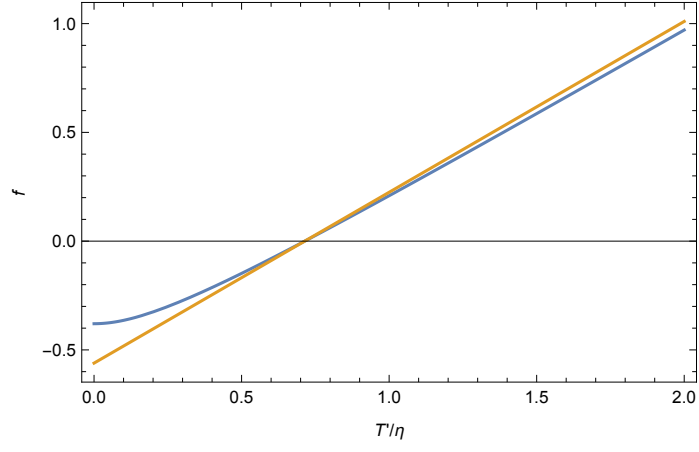}
\caption{\label{forcefig} Force between a nonreciprocal  nanoparticle and a
metal plate with finite conductivity out of thermal equilibrium.  The 
temperature of both 
the plate and the background electromagnetic field is fixed
at room temperature, which corresponds to $T/\eta=0.714$ for gold.  The 
polarizability of the nanoparticle 
is described by the model (\ref{modela}). 
The temperature of the 
nanoparticle, $T'$, is given in units of the damping parameter for gold,
$\eta$.  For comparison, the straight line shows the force when both
temperatures are large.}
\end{figure}

Now for the nonreciprocal polarizability, let us use the model (\ref{modela}),
where we now assume that the magnetic field (confined to the particle)
lies in the $y$ direction.
This leads directly to the following formula for the force, ($x=\omega/\eta$)
\be
F_x=\frac{3 V}{4\pi^2a^4}\frac\nu\eta \omega_c f(\beta\eta,\beta'\eta),
\quad f(\beta\eta,\beta'\eta)=-\int_0^\infty dx\frac{x}{(x^2+1)^2}
\left[\frac1{e^{\beta\eta x}-1}-\frac1{e^{\beta'\eta x}-1}\right],
\ee
where the integral follows from Appendix \ref{appa},
\be
\int_0^\infty dx\frac{x}{(x^2+1)^2}\frac1{e^{\beta\eta x}-1}
=\frac14\left[1+\frac{\pi}{\beta\eta}-\frac{\beta\eta}{2\pi}\psi'\left(
\frac{\beta\eta}{2\pi}\right)\right].
\ee
The dimensionless force $f$ 
 is plotted in Fig.~\ref{forcefig}, and compared to the high-temperature
approximation. The prefactor, for $\nu=\eta$, $a=1\,\mu$m, and 
the radius of the nanosphere being $100$ nm, is $5\times 10^{-21}$ N.

%If $\xi(\omega)$ falls off fast enough as $\omega\to \infty$, we can do
%a high-temperature expansion, and write the integral appearing here as
%\be
%(T-T')\int_0^\infty d\omega\,\xi(\omega)=\frac\pi{\omega_0}(T-T'),\label{hit}
%\ee if, as an example, we choose 
%\begin{equation}
%\xi(\omega)=1/(\omega^2+\omega_0^2).\label{model}
%\end{equation}   On the
%other hand, if $\xi(\omega)=1$, the frequency integral in 
%Eq.~(\ref{lowfreqf}) is 
%\be
%\frac{\pi^2}3(T^2-T^{\prime 2}).\label{lt}
%\ee
%Of course, because of the low-frequency approximation, this can only
%be valid for low temperatures.  The approximation (\ref{hit}) is compared
%with the formula (\ref{lowfreqf}) for the model (\ref{model})
%in Fig.~\ref{fig1}.

%\begin{figure}
%\includegraphics{lowf.eps}
%\caption{The function $F_y$ (apart from the constant prefactors
%in the second line of Eq.~(\ref{lowfreqf})) in the
%low-frequency approximation.  Here we take $\omega_0T'=1$, and plot the force
%as a function of $\omega_0T$. The dotted red line is the linear
%approximation (\ref{hit}).}
%\label{fig1}
%\end{figure}
  
\subsection{Transverse force in presence of a perfectly conducting plate}
\label{sec:trfpc}
\begin{figure}
\includegraphics{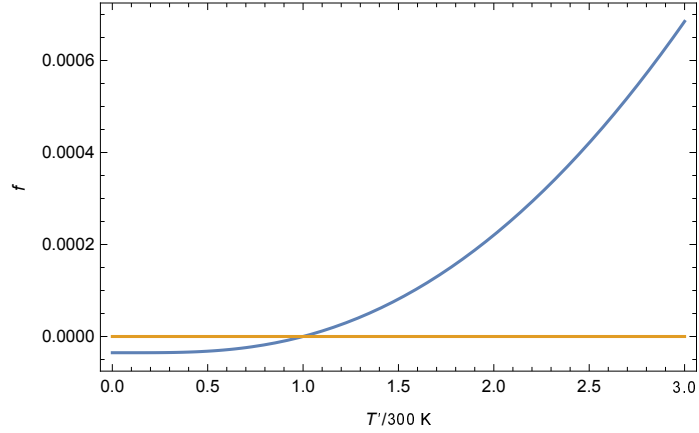}
\caption{\label{transffig} The transverse force
 given in Eq.~(\ref{trf})
as a function of the 
temperature of the nonreciprocal nanoparticle relative to room temperature,
300 K, for the environment and perfectly conducting  plate at room temperature.
Here we take the separation $a$ of the nanoparticle and the plate to be 100 nm,
and the damping to be that appropriate for gold, 0.035 eV.  In this
case, the prefactor in the force in Eq.~(\ref{trf}),
for a gold nanosphere of 10 nm radius is 
$1.2\times 10^{-20}$ N, so this would be challenging to observe.}  
\end{figure}

If the slab is a perfect conductor, with $r^H=1$, the formula for the
transverse force simplifies considerably.  The imaginary part of
the Green's function then requires that $\omega^2>k^2$, and so the integral
over the wavenumber is
\be
\int \frac{(d\mathbf{k}_\perp)}{(2\pi)^2}k_x^2 \im e^{-2\kappa a}
=-\frac1{4\pi}\frac1{(2a)^4}[6u\cos u+2(u^2-3)\sin u],
\ee
where $u=2\omega a$.  Then, the transverse force is
\be
F_x=\frac{\omega_c\eta\omega_p^2 V}{2\pi^2 a}f_0(\epsilon,b,b'),\quad
f_0(\epsilon,b,b')=\int_0^\infty du
\frac{6u\cos u+2(u^2-3)\sin u}{(u^2+\epsilon^2)^2}\left[
\frac1{e^{ u b}-1}-\frac1{e^{u b'}-1}\right],\label{trf}
\ee
where $\epsilon=2\eta a$, $b=1/(2aT)$, and $b'=1/(2aT')$.
The integral $f$ is plotted in Fig.~\ref{transffig}
as a function of the nanoparticle temperature $T'$
for the environment at room temperature, for a separation of $a=100$ nm, with
a damping parameter appropriate for gold, $\eta=0.035$ eV.  Note, that the
high-temperature limit for the force is given by  $f_0\sim \frac\pi{8 b'}$,
for $b'\ll 1/\epsilon, b$, but this limit requires very high temperatures which
are not accessible in practice.
It is noteworthy that Figs.~\ref{forcefig} and \ref{transffig} are 
qualitatively (but not quantitatively) similar,
given that the physical mechanisms invoked are rather different.

It is easily seen that the lateral force rapidly vanishes as $a\to
\infty$, consistent with the absence of a quantum vacuum force.

\begin{figure}
\includegraphics{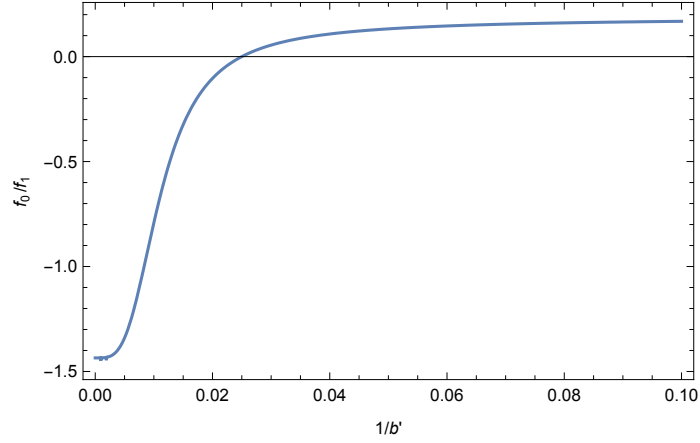}
\caption{\label{fig:termvel}
Terminal velocity of a nonreciprocal nanoparticle near a perfectly
conducting plate in units of $2\omega_c a$.  Here it is assumed that
the plate and the background are at temperature $T=300$ K, that the particle
is made of gold, and it is $a= 100$ nm above the surface of the plate.  The 
temperature of the nanoparticle is $T'=1/(2ab')$.  Thus, the highest particle
temperature displayed on the graph is 1200 K.}
\end{figure}

We expect, as we saw for the torque, that this force will be resisted by the
quantum vacuum friction in the presence of the plate, which for low velocities
will lead to a terminal velocity, according to
\be
m\frac{dv}{dt}=F_0-vF_1' \Rightarrow v(t)=\frac{F_0}{F_1'}\left(1-e^{-F_1' t/m}
\right).
\ee
We require, then, the nonequilibrium frictional force in the presence of a
conducting plate, which we derive in Appendix~\ref{appb}.  Using the same
model for the permittivity of the nanoparticle, the linear term in the
friction is
\be
F_1'=\frac{\omega_p^2\eta V}{\pi^2(2a)^2}f_1(\epsilon,b,b'),
\ee
where
\bea
f_1(\epsilon,b,b')&=&\int_0^\infty du\frac{u^3}{u^2+\epsilon^2}\bigg\{
\left[1-\frac{2\cos u+(u^2-2)\sin u}{u^3}\right]
\left[\frac1{e^{b'\!u}-1}-\frac1{e^{bu}-1}\right]\nonumber\\
&&\quad\mbox{}+\frac1{12}\frac{bu}{\sinh^2(bu/2)}\left[1-3
\frac{-u(u^2-12)\cos u+(5u^2-12)\sin u}{u^5}\right]\bigg\}.\label{transfric}
\eea
Indeed, $f_1$ is always positive, corresponding to a frictional drag, and the
corresponding terminal velocity is
\be
v_T=\frac{F_0}{F_1'}=2\omega_c a \frac{f_0}{f_1}.\label{term2}
\ee
The scale factor here is  small compared to the speed of light: 
for a particle 100 nm above
the plate, $2 \omega_c a = 10^{-4}$.  The ratio of $f_0/f_1$ is shown in
Fig.~\ref{fig:termvel}.
The apparent saturation of the terminal velocity near 0.2 is illusory; for 
still larger temperatures, the terminal velocity tends to zero, since the
frictional force rapidly increases with temperature.
However, for these nominal values, the damping time is  long:
\be
t_0=\frac{m}{F_1'}\sim 2\times 10^3\,\mbox{s}\frac1{f_1}\sim 10^6 \,\mbox{s},
\label{relax2}
\ee 
if the particle is at twice room temperature.

\section{Relaxation to Thermal Equilibrium}
\label{sec:cool}
All of the above considerations assumed that the temperatures of the body
and of the background are constant.  Of course, this will not be so unless
some mechanism keeps the system out of thermal
equilibrium.  Here, we will calculate
the time it would take for such a body 
at rest to come to thermal  equilibrium
with its environment.  We cannot regard the body to be
 a black body, but we
can calculate the rate at which it loses heat from the power
(for an isotropic body)\footnote{For the purpose of the rough
calculation  presented  here, we will ignore the small nonreciprocal effects.}
 \cite{guo2} 
\be
P(T,T')=\frac1{\pi^2}\int_0^\infty d\omega\,\omega^4 \im\alpha(\omega)
\left[\frac1{e^{\beta\omega}-1}-\frac1{e^{\beta'\omega}-1}\right]
=\frac{dQ}{dt}.
\label{powerloss}
\ee
This is related to the rate of change of temperature of the 
body by its heat capacity:
\be
\frac{dQ}{dt}=C_V(T')\frac{dT'}{dt}.
\ee
Thus, the time it takes for the body 
to cool from temperature $T'_0$ to
temperature $T'_1$, where $T'_0>T'_1>T$, is 
\be
t=\int_{T'_0}^{T'_1} dT'\frac{C_V(T')}{P(T,T')}.\label{ct}
\ee

To proceed, we need a model for the  heat capacity of the body.  Such
is provided by the Debye model,\footnote{We consider bulk effects only,
and ignore surface effects, for the purpose of a rough estimate.}
 which is satisfactory for simple crystals
 (see Ref.~\cite{LL}):
\be
C_V(T)=9N\!\left(\frac{T}\Theta\right)^{\!3}
\!\!\int_0^{\Theta/T}\!\! dx\frac{x^4 e^x}{(e^x-1)^2}.
%3N\left[D\left(\frac\Theta{T}\right)-\frac\Theta{T}D'\left(\frac\Theta{T}
%\right)\right],\quad D(x)=\frac3{x^3}\int_0^x dy\frac{y^3}{e^y-1},
\ee
where $N$ is the number of atoms constituting the body,
and $\Theta$ is the Debye temperature.
This interpolates between the low- and high-temperature limits:
\be
C_V(T)\sim 3N\left\{\begin{array}{cc}
1,&T\gg\Theta,\\
\frac{4\pi^4}5\!\!\left(\frac{T}{\Theta}\right)^{\!3},
&T\ll\Theta.\end{array}\right.
\label{Debyelimits}
\ee
Since the Debye temperature for gold is about $\Theta=170$ K, the 
high-temperature approximation would seem appropriate for an estimate at
room temperature and above.

We finally need a model for the imaginary part of the polarizability of
the body.  The Lorenz-Lorentz model would give
\be
\im \alpha(\omega)=\frac{V\omega_p^2\omega\eta}{(\omega_1^2-\omega^2)^2+
\omega^2\eta^2}\approx \frac{V\omega_p^2\omega\eta}{\omega_1^4},
\label{fig:LLmodelcool}
\ee
where, for a metal (Drude model), $\omega_1=\omega_p/\sqrt{3}$.  The 
approximation here is appropriate if, as expected, $\omega_1\gg \omega, \eta$.
Inserting this approximation into the formula (\ref{powerloss}) we obtain
\be
P(T,T')\approx \frac{8 \pi^4}7 \frac{V\eta}{\omega_p^2}(T^6-T^{\prime6}).
\ee

\begin{figure}
\centering
\begin{minipage}{.45\linewidth}
\includegraphics[width=1.2\linewidth]{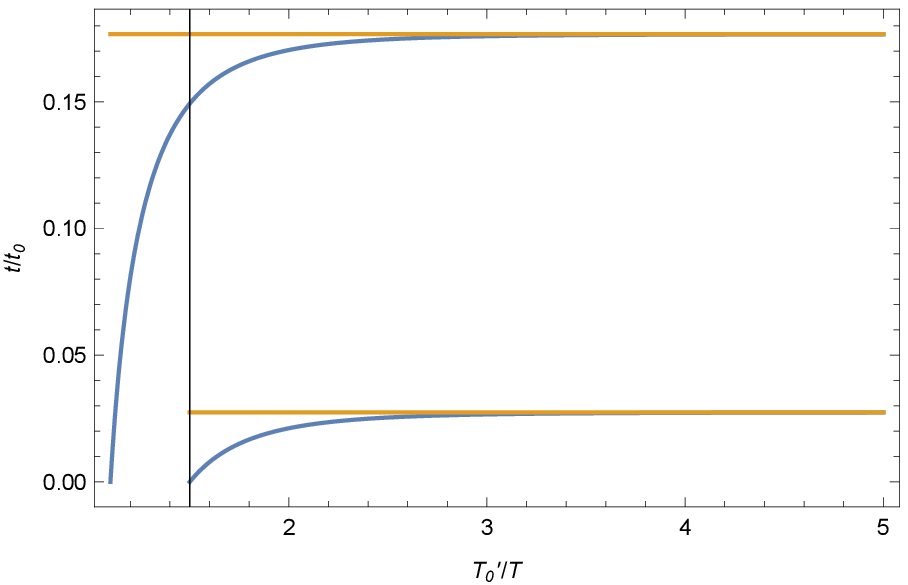}
%\caption*{Cooling time $t/t_0$ for enviromental temperature $T=300$ K}
%\label{fig:cool}
\end{minipage}%
\hfill
\begin{minipage}{.45\textwidth}
\includegraphics[width=1.2\linewidth]{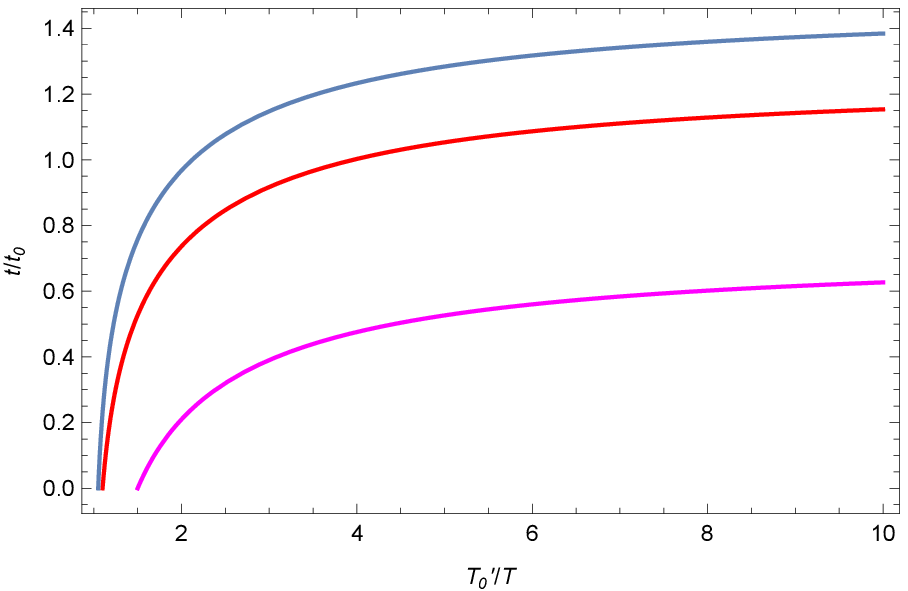}
%\caption*{Cooling time $t/t_0$ for enviromental temperature $T=1$ K}          %\label{fig:coolb}
\end{minipage}
\caption{Time required for a body to cool from temperature
$T'_0$ to temperature $T'_1$ for different environmental temperatures $T$. Here
$T'_0>T'_1>T$.  The left panel is for the environmental temperature $T=300$ K,
and the right is for $T=1$ K.  On the left,
the upper set of curves is for $T'_1/T=1.1$, and the lower curves
are for $T'_1/T=1.5$.  On the right, the three curves from top to bottom
are for $T'_1/T=1.05$, 1.1, and 1.5.
The times are scaled by the prefactor, $t_0$, in 
Eq.~(\ref{cooltime}), which for a gold body 
 evaluates to about $10^4$~s, for the environment
at room temperature, and  by $\tilde t_0$ (Eq.~(\ref{ratts})), 
which is about $10^{11}$~s, for $T=1$ K.}
\label{fig:tcool}
\end{figure}

\begin{figure}
\includegraphics{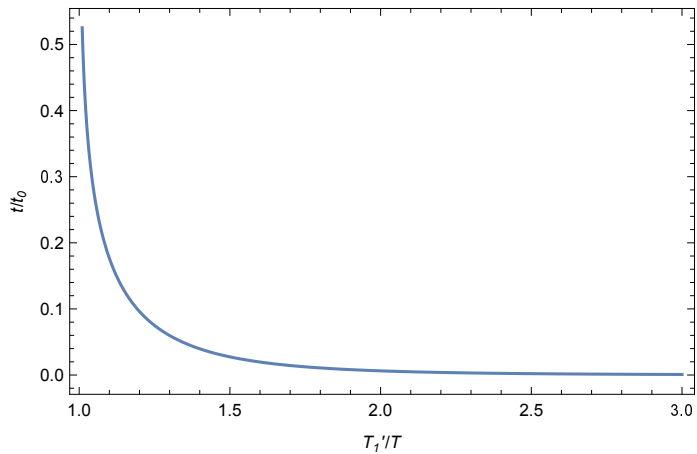}
\caption{\label{fig:tcool2} 
The cooling time, in terms of the scale $t_0$, for the
nanoparticle to cool from a high temperature to $T_1$.  It takes an 
increasingly long time to get very close to the environmental temperature.} 
\end{figure}

Now we compute the cooling time from Eq.~(\ref{ct}):
\be
t= t_0\int_{T'_0/T}^{T'_1/T}\! du\frac1{1-u^6},\quad T'_0>T'_1>T, \quad \mbox{where}
\quad
t_0=\frac{21}{8\pi^4}n\frac{\omega_p^2}\eta\frac1{T^5}\label{cooltime}
\ee
Here, $n$ is the number density of atoms in the body.
The relaxation scale, $t_0$, is independent of the volume of the particle,
and is about $10^4$ seconds for gold, for an environmental temperature of
300 K.
The cooling time
 diverges as $T'_1\to T$, but cooling to a temperature slightly above
the environmental temperature takes a finite time.  The integral here is
elementary, but the resulting expression is not very 
illuminating.  We content ourselves
by showing some representative values in left panel of
Fig.~\ref{fig:tcool}.
It will be seen that if $T'_0$ is appreciably larger than $T$ the cooling
time rapidly saturates to an asymptotic value. If we then take $T'_0$ to be
large, 
Fig.~\ref{fig:tcool2} shows how long it will take to reach a multiple of
the environmental temperature.
Thus, we see that the terminal angular velocity seen in Eq.~(\ref{term1}) 
and the terminal linear velocity obtained in 
Eq.~(\ref{term2}) will not be achievable unless some mechanism %(which
%already must exist to acoount for the nonreciprocity) 
maintains the
thermal imbalance, because the time scales for achieving those velocities,
Eqs.~(\ref{relax1}) and (\ref{relax2}), are much longer than the cooling
time found here.

If the environmental temperature is very low, $T\ll\Theta$, the cooling time
is very much longer.  The analysis proceeds as above, using
the low temperature limit in Eq.~(\ref{Debyelimits}), 
with the result for
the cooling time being
\be
t=\tilde{t}_0\int_{(T'_0/T)^2}^{(T'_1/T)^2}\!\! dy\frac{y}{1-y^3}, \quad
\tilde{t}_0=\frac{21}{20}n\frac{\omega_p^2}\eta
\left(\frac{T}{\Theta}\right)^{\!3}\!\frac1{T^5}.
\ee
The integral, which has a relatively simple analytic form, is shown in
the right panel of Fig.~\ref{fig:tcool}.
The ratio of the time scales in the two cases is
\be
\frac{\tilde{t}_0}{t_0}=\frac{2\pi^4}{5}\left(
\frac{T_{\rm low}}{\Theta}\right)^{\!\!3}
\left(\frac{T_{\rm high}}{T_{\rm low}}\right)^{\!\!5}\sim 10^7,\label{ratts}
\ee
for $T_{\rm low}=1$ K, $T_{\rm high}=300$ K, and $\Theta=170$ K for gold,
so terminal velocities might be achievable.

\section{Lorenz-Lorentz correction}
\label{sec:LL}
Hithertofore, we have ignored the Lorenz-Lorentz correction familiar in
passing from the permittivity of a body to its polarizability.  
This was because the forces and torques were derived directly from the 
macroscopic susceptibilities appropriate for a dissipative metal body.
In the case of small bodies, we could always pass from the susceptibility
to the mean polarizability by integrating over the volume of the body.
However, as is evident from the discussion in the preceding section [see
Eq.~(\ref{fig:LLmodelcool})], the effect
of the medium on the local electric field can result in a large correction
in the case of metal bodies.  

The difficulty is that the simple Lorenz-Lorentz model is ordinarily derived
using spherical symmetry.  The relation between the electric polarizability
and the  permittivity is, in the isotropic case, in HL units,
\be
\alpha=\frac{\varepsilon-1}{\varepsilon+2}4\pi a^3.
\ee
This is not valid for a nonsymmetric permittivity;
for example, see Ref.~\cite{Pal}.
However, for $\omega_c$ small, the nonsymmetric nature is small, so,
for the purpose of an estimate, we use, as in Refs.~\cite{fan,strekha}, 
the matrix generalization of the above:
\be
\bm{\alpha}=(\bm{\varepsilon-1})(\bm{\varepsilon+2})^{-1}4\pi a^3.
\ee
It is quite straightforward to compute the components of this matrix: the
term we need for the torque in Eq.~(\ref{vact}) is, for $\omega_p\gg \omega
\sim T$, 
\be
\re\alpha_{xy}\approx54 V \frac{\omega^2\omega_c \eta}{\omega_p^4}.
\ee
When this is inserted into Eq.~(\ref{vact}), we obtain
\be
\tau_z=\frac{32}7\pi^4 V
\frac{\omega_c\eta}{\omega_p^4} T^6\left[1-\left(\frac{T'}{T}\right)^{\!\!6}
\right].
\ee
Putting in the numbers for a 100 nm gold nanosphere, the coefficient of 
$(1-T^{\prime 6}/T^6)$ is about $5\times 10^{-36}$ Nm, 
some 11 orders of magnitude
smaller than that found at the end of Sec.~\ref{vactorque}.

We can also repeat the calculation of the terminal angular velocity in 
Sec.~\ref{sec:rot} in this Lorenz-Lorentz model.  Then,
\be
\tau'_1=-\frac1{6\pi^2}\left(3+\beta\frac\partial{\partial\beta}\right)
\int_0^\infty
d\omega\,\omega^2\im(\alpha_{xx}+\alpha_{yy})(\omega)\left(
\frac1{e^{\beta'\!\omega}-1}
-\frac{1}{e^{\beta\omega}-1}\right),
\ee
where in our model
\be
\im(\alpha_{xx}+\alpha_{yy})\approx 18 V \frac{\omega \eta}{\omega_p^2},
\ee
which implies
\be
\tau_1'=\frac{2\pi^2}5\frac{V\eta}{\omega_p^2}T^4\left[1+3\left(\frac{T'}T
\right)^{\!\!4}\right].
\ee
Note that $\tau_1'$ is always positive, indicating that it always 
opposes the rotation.
The corresponding terminal angular velocity is
\be
\Omega_T=\frac{\tau_0}{\tau_1'}=\frac{80}7 \pi^2\frac{\omega_c}{\omega_p^2}T^2
\frac{1-\frac{T^{\prime6}}{T^6}}{1+3\frac{T^{\prime 4}}{T^4}},
\ee where the prefactor, independent of $T'$, 
implies a substantial angular velocity for
gold at room temperature: $\sim 10^8$ s$^{-1}$.
Although the time required to reach such a velocity is very long,
$t_0=I/\tau_1'\sim 10^{13}$ s,   the initial
angular acceleration is not so small,
\be
\dot\Omega(0)=\frac{\Omega_T}{t_0}\sim 10^{-5}\,\mbox{s}^{-2}.
\ee
While this angular acceleration is
 10 orders of magnitude smaller than that found
without the Lorenz-Lorentz correction in Eq.~(\ref{Omega0}), 
the body will acquire a measurable angular velocity after a relatively
small period of observation.

However, is this correction valid or even necessary for a metal nanoparticle?
The discussion in Secs.~\ref{vactorque}--\ref{transverseforce} 
is based on describing the susceptiblity of a metal by
the phenomenological Drude
model, which should include, approximately, all internal effects.
There is a large literature on the subject of ordinary polarizabilities
of metal nanoparticles---see Refs.~\cite{Puska,Kheirandish}, 
for example---where it is seen that both classical and approximate
quantum mechanical treatments are inadequate. We are unaware of comparable
work in the nonreciprocal case.  So, to some extent, 
 the issue of applying the Lorenz-Lorentz correction  remains open.
In this paper we are interested in the interaction between the 
electromagnetic field
and the body, the electromagnetic properties of which are specified by
a given susceptibility, so that the crude models for the latter should only
 be taken as illustrative.
 
\section{Conclusions}
In this paper we have concentrated on analysis to first order in the
susceptibility, to better understand the effects of nonreciprocal materials
on torque and on forces for bodies out of thermal equilibrium with
their environment.  Time-reversal
symmetry is broken by these materials, so spontaneous forces and torques
are possible.  Of course, time reversal symmetry is not broken by 
electrodynamics, whether classical or quantum; rather the nonreciprocity is a
consequence of an external agent, such as a magnetic field, that is
encoded in the dielectric response of the materials.

Interestingly, potentially observable phenomena are, nevertheless, predicted.
A nonreciprocal body out of thermal
 equilibrium will spontaneously start to rotate,
and reach a substantial terminal angular velocity.  Such a body will not feel
a net force to first order in the susceptibility.  However, if the body
is placed near a  translationally invariant
surface, even a perfect conductor, then a force parallel to the surface
would arise. The presence of such a surface would tend to
suppress the vacuum torque.  A potentially observable terminal 
linear velocity
arises here as well, although the time scales are such that it would
be difficult to keep the system out of thermal  equilibrium.
A possible drastic reduction in the strength of these
nonreciprocal effects, due to the Lorenz-Lorentz correction for dielectric
susceptibilities, is discussed in the penultimate section, 
although it seems the angular and linear accelerations
might still  be amenable to observation.

Elsewhere, we will examine higher-order effects, to see how phenomena such as
vacuum self-propulsion can arise, even for a reciprocal body 
\cite{Reid,muller}.

\begin{acknowledgments}
The work of KAM and XG was supported in part by a grant from the US
National Science Foundation, No.~2008417.
The work of GK and KAM was supported in part by a grant from the US
National Science Foundation, No.~PHY-1748958.  In particular,
GK and KAM thank KITP for its hospitality, and
Benjamin Strekha for conversations there, which stimulated
the research carried out here.
We thank Steve Fulling, Li Yang, Prachi Parashar,  Shadi Rezaei,  and 
Venkat Abhignan for collaborative assistance.  %We thank Benjamin Strekha
%for helpful conversations.  GK and KAM thank the KITP for hospitality,
%and NSF grant No.~PHY-1748958?.  
This paper reflects solely the authors'
personal opinions and does not represent the opinions of the authors'
employers, present and past, in any way.
For the purpose of open access, the authors have applied a CC BY public 
copyright license to any Author Accepted Manuscript version arising from 
this submission.

\end{acknowledgments}

\appendix

\section{Evaluation and expansion of integrals}
\label{appa}

Here, we express the integrals encountered in the text in terms of the digamma 
and trigamma functions, $\psi(z)$ and $\psi'(z)$, and provide corresponding 
low- and high-temperature expansions.

 Differentiation of Binet's second integral representation of the log gamma 
function immediately yields the integral representation
\begin{equation}
\psi(z)=\ln z -\frac{1}{2z}-2\int_0^{\infty}dx\,\frac{x}{x^2+1} 
\frac{1}{e^{2\pi z x}-1}.\label{A1}
\end{equation}
Thus, 
\begin{equation}
I_1(\beta\eta)\equiv \int_0^{\infty}dx\,\frac{x}{x^2+1}
\frac{1}{e^{\beta\eta x}-1}=\frac12\left[-\frac{\pi}{\beta\eta}
+\ln\left(\frac{\beta\eta}{2\pi}\right)-\psi\left(\frac{\beta\eta}{2\pi}
\right)\right].\label{A2}
\end{equation}

 Since
\begin{equation}
\beta\frac{\partial}{\partial \beta} \frac{1}{e^{\beta\eta x}-1}=\eta
\frac{\partial}{\partial \eta} \frac{1}{e^{\beta\eta x}-1}
=x\frac{\partial}{\partial x} \frac{1}{e^{\beta\eta x}-1},\label{A3}
\end{equation}
it follows that
\begin{equation}
\beta\frac{\partial}{\partial \beta} I_1(\beta\eta)=\eta
\frac{\partial}{\partial \eta} I_1(\beta\eta)=\int_0^{\infty}dx\,
\frac{x^2}{x^2+1}\frac{\partial}{\partial x}\frac{1}{e^{\beta\eta x}
-1}=2\int_0^{\infty}dx\,\left[\frac{x^3}{(x^2+1)^2}-\frac{x}{x^2+1}\right]
\frac{1}{e^{\beta\eta x}-1}.\label{A4}
\end{equation}
Thus,
\begin{equation}
I_2(\beta\eta)\equiv \int_0^{\infty}dx\, \frac{x^3}{(x^2+1)^2}
\frac{1}{e^{\beta\eta x}-1}=\left[1+\frac{\beta}{2}
\frac{\partial}{\partial\beta}\right]I_1(\beta\eta)=\left[1+\frac{\eta}{2}
\frac{\partial}{\partial\eta}\right]I_1(\beta\eta).\label{A5}
\end{equation}
Hence, from Eq.~(\ref{A2}),
\begin{equation}
I_2(\beta\eta)= \frac14\left[-\frac{\pi}{\beta\eta}+2\ln\left(
\frac{\beta\eta}{2\pi}\right)+1-2\psi\left(\frac{\beta\eta}{2\pi}\right)
-\frac{\beta\eta}{2\pi}\psi'\left(\frac{\beta\eta}{2\pi}\right)\right].
\label{A6}
\end{equation}

 Using the series representation
\begin{equation}
\psi(z)=-\frac{1}{z}-\gamma_E+z\sum_{k=1}^{\infty}\frac{1}{k(z+k)}, \label{A7}
\end{equation}
where $\gamma_E$ is the Euler-Mascheroni constant, we readily obtain from 
Eqs.~(\ref{A2}) and (\ref{A6}) the small $\beta\eta$ (or high-temperature) 
expansions
\begin{subequations}
\begin{equation}
I_1(\beta\eta) \sim \frac12\left[\frac{\pi}{\beta\eta}+\ln\left(
\frac{\beta\eta}{2\pi}\right)+\gamma_E\right] \quad\quad (\beta\eta \to 
0)\label{A8}
\end{equation}
and
\begin{equation}
I_2(\beta\eta)\sim \frac14\left[\frac{\pi}{\beta\eta}+2\ln\left(
\frac{\beta\eta}{2\pi}\right)+1+2\gamma_E\right] \quad \quad (\beta\eta\to 0).
\label{A9}
\end{equation}
\end{subequations}

 Likewise, using the asymptotic representation
\begin{equation}
\psi(z) \sim \ln z -\frac{1}{2z}-\sum_{k=1}^{\infty}
\frac{B_{2k}}{2kz^{2k}} \quad \quad (z\to \infty),\label{A10} 
\end{equation}
there follow from Eqs.~(\ref{A2}) and (\ref{A6}) the large 
$\beta\eta$ (or low-temperature) expansions
\begin{subequations}
\begin{equation}
I_1(\beta\eta) \sim \frac{\pi^2}{6\beta^2\eta^2}
-\frac{\pi^4}{15\beta^4\eta^4}\quad\quad (\beta\eta \to \infty)\label{A11} 
\end{equation}
and
\begin{equation}
I_2(\beta\eta) \sim \frac{\pi^4}{15\beta^4\eta^4} \quad\quad 
(\beta\eta\to \infty).\label{A12}
\end{equation}
\end{subequations}

\section{Out-of-equilibrium frictional force near perfectly conducting plate}
\label{appb}
Following the discussion in Ref.~\cite{guo2}, it is easy to derive the
expression for the frictional force used in Sec.~\ref{sec:trfpc}.  The general
expression for the force is
\be
F=\int\frac{d\omega}{2\pi}\frac{d\mathbf{k}_\perp}{(2\pi)^2}(k_x+\omega v)
\tr \Im\bm{\alpha}(\omega)\Im \mathbf{g'}(\omega,\mathbf{k}_\perp)
\left[\coth\frac{\beta\gamma(\omega+k_x v)}2-\coth\frac{\beta'\omega}2
\right].\label{genf}
\ee
Here $\mathbf{g'}$  is the reduced Green's function in the rest frame
of the particle.  For a perfectly conductimg plate, as with the vacuum,
$\mathbf{g'}=\mathbf{g}$, the Green's function in the frame of the plate
and vacuum.  
From this we can calculate both the frictional force and the propulsive
force.  The latter is present at $v=0$, and arises from the antisymmetric
parts of the polarizability and the Green's dyadic, and is given in 
Sec.~\ref{transverseforce}.  The diagonal parts 
of both these tensors correspond
to friction.  According to the model (\ref{modforchi}) 
with $\omega_c$ neglected,
the diagonal terms of the imaginary part of the  polarizability are all 
equal to
\be
\im\alpha_d=\frac{V\omega_p^2\omega\eta}{(\omega_0^2-\omega^2)^2+\omega^2
\eta^2}.
\ee
That leaves us with the imaginary part of the trace of $\mathbf{g}$:
\be
\im\tr\mathbf{g}=\sgn(\omega)\left[\frac{\omega^2}{\sqrt{\omega^2-k^2}}
-\sqrt{\omega^2-k^2}\cos\left(2a\sqrt{\omega^2-k^2}\right)\right]
\theta(\omega^2-k^2).
\ee

Now when we expand Eq.~(\ref{genf}) to first order in $v$ we obtain two
terms:
\be
F=F^{(1)}+F^{(2)}.
\ee
Here the first term comes from expanding the hyperbolic cotangent:
\be
F^{(1)}=-\frac{\beta v}{8\pi^2}\int_0^\infty d\omega\im\alpha_d(\omega)
\frac{\omega^5}{\sinh^2(\beta\omega/2)}\left\{\frac23-\frac2{x^5}\left[
-x(x^2-12)\cos x+(5x^2-12)\sin x\right]\right\},
\ee
where we've carried out the elementary integration over $\mathbf{k}_\perp$.
Note that the first term here corresponds to the usual Einstein-Hopf effect.
The $F^{(2)}$ contribution to the force
corresponds to the $\omega v$ prefactor in Eq.~(\ref{genf}),
and is a non-equilibrium friction contribution:
\be
F^{(2)}=\frac{v}{2\pi^2}\int_0^\infty d\omega\im\alpha_d(\omega)\omega^4
\left(\coth\frac{\beta\omega}2-\coth\frac{\beta'\omega}2\right)\left[
1-\frac1{x^3}\left(2x\cos x+(x^2-2)\sin x\right)\right],
\ee
again after carrying out the wavenumber integration.
The sum of these two terms is Eq.~(\ref{transfric}) if we set $\omega_0=0$,
appropriate for a metal nanoparticle.

\end{document}